\documentclass[conference]{IEEEtran}
\IEEEoverridecommandlockouts
\usepackage{cite}
\usepackage{amsmath,amssymb,amsfonts}
\usepackage{algorithmic}
\usepackage{graphicx}
\usepackage{textcomp}
\usepackage{xcolor}

\usepackage{tabularx}
\usepackage{pifont}
\usepackage{multirow}

\usepackage{subcaption}
\usepackage{indentfirst}

\usepackage{capt-of}
\usepackage{varwidth}

\def\BibTeX{{\rm B\kern-.05em{\sc i\kern-.025em b}\kern-.08em
    T\kern-.1667em\lower.7ex\hbox{E}\kern-.125emX}}
\begin{document}

\title{Prediction-driven resource provisioning for serverless container runtimes}

\author{
\IEEEauthorblockN{Dimitrios Tomaras, Michail Tsenos, Vana Kalogeraki}
\IEEEauthorblockA{Department of Informatics\\
Athens University of Economics and Business, Athens, Greece\\
\{tomaras, tsemike, vana\}@aueb.gr}
}

\maketitle

\begin{abstract}
In recent years Serverless Computing has emerged as a compelling cloud based model for the development of a wide range of data-intensive applications. 
However, rapid container provisioning introduces non-trivial challenges for FaaS cloud providers, as (i) real-world FaaS workloads may exhibit highly dynamic request patterns, (ii) applications have service-level objectives (SLOs) that must be met, and (iii) container provisioning can be a costly process. In this paper, we present SLOPE, a prediction framework for serverless FaaS platforms to address the aforementioned challenges. Specifically, it trains a neural network model that utilizes knowledge from past runs in order to estimate the number of instances required to satisfy the invocation rate requirements of the serverless applications. In cases that a priori knowledge is not available, SLOPE makes predictions using a graph edit distance approach to capture the similarities among serverless applications. Our experimental results illustrate the efficiency and benefits of our approach, which can reduce the operating costs by 66.25\% on average.
\end{abstract}


\section{Introduction}

Serverless computing, and in particular Function as a Service (FaaS), is becoming an increasingly popular cloud programming model\cite{tsenos2023acsos,peri2023acsos}, fueled by the recent demand to host services on provisioned cluster infrastructures and the paradigm shift towards interconnected IoT applications, devices and platforms \cite{tomaras2018modeling,zacheilas2018orion,dedousis2018fly}. It offers an intuitive, event-based interface for developing cloud-based applications, that makes the writing and deployment of scalable microservices easier and cost effective. 
This computing model has additional advantages including lower operational and deployment costs due to its unique pricing policy (based on a pay-as-you-use model) where users do not explicitly provision or configure virtual machines (VMs) or containers but
they only get charged based on the number of resources consumed by the application functions during execution\cite{elgamal2018costless,rajan2018serverless}.
The serverless computing model has been successfully adopted
in a wide range of application domains, including, processing event streams, next-generation web services 
and applications\cite{tomaras2022practical,tsenos2022amesos}, etc.
\footnote{https://docs.microsoft.com/en-us/dotnet/architecture/serverless/serverless-design-examples}
All major commercial cloud service providers are now offering serverless computing platforms,
including AWS Lambda (https://aws.amazon.com/lambda/), Google Cloud Functions (https://cloud.google.com/functions) and Azure Functions (https://azure.microsoft.com/en-gb/products/functions/).

The primary reason that makes the FaaS model so appealing is the fundamental resource elasticity it provides; a FaaS platform allows user applications to scale up to tens of thousands of cloud functions on demand, in seconds, with no advance
notice. Recent research has shown, that, serverless architectures can be successfully 
utilized to support the execution of data intensive applications and workloads with high demand for data parallelism\cite{jiang2021towards}. For instance, big data analytics tasks exploit the inherent parallelism of the serverless architectures
where a number of stateless functions are launched, each processing a different batch of data, without managing or maintaining any servers. 

So what makes resource provisioning for serverless applications an essential step? 
{\it First,} custom container images are large in sizes (larger than 1.3 GB \cite{wang2021faasnet,zhao2019large}). 
Fetching such large container images from
a remote registry can incur significant cold startup latencies,
which can be up to several minutes. 
Furthermore, executing a serverless function requires the function code (e.g., user code, language runtime libraries) to be brought from persistent storage into main memory (a phenomenon known as {\it cold-start}). Keeping the functions in memory at all times may be prohibitively expensive for the service provider, 
as function calls can be very sparse or other times highly dynamic. 
{\it Cold start} refers to the set-up time required by the FaaS provider to get a serverless function’s environment up and running before executing the function. 
The cold start time can be a significant fraction of the function's execution time and rises sharply with an increased but unpredictable number of function requests \cite{liu2023faaslight,agarwal2021reinforcement,silva2020prebaking,wang2021faasnet,oakes2018sock,du2020catalyzer,shillaker2020faasm,shahrad2020serverless,fuerst2021faascache,hunhoff2020proactive}.
Furthermore, the functions can vary widely with respect to their resource needs and invocation frequencies from multiple triggers, making the prediction of function invocations a rather challenging problem. 
As a result, the high cost of the container startup process makes it extremely challenging for FaaS providers to deliver high elasticity services. 

To address the cold-start problem in serverless functions recent research 
has generated different solution approaches, divided into 
two main areas: i) container optimization \cite{silva2020prebaking,hunhoff2020proactive,oakes2018sock,du2020catalyzer,shillaker2020faasm} 
and ii) prediction methods \cite{shahrad2020serverless,fuerst2021faascache,fu2019edgewise}. Container optimization methodologies focus on enhancing the creation process of containers, such as maintaining pools of ready containers or optimizing how the building blocks of a container are being loaded\cite{oakes2018sock}. On the other hand, prediction methods focus on estimating the number of resources based only on predictable behaviours
or certain conditions
\cite{shahrad2020serverless,fuerst2021faascache}. 
However, both approaches are inadequate since 
we would often need to adjust dynamically the trained model, e.g., due to fluctuating, highly dynamic or bursty user request patterns, or we would need to use extra resources and have a set of containers \textit{prewarmed} to start execution.

{\it Second,} existing approaches, such as schedulers designed for VM
placement or web load balancers, are not well suited for scheduling 
the execution of serverless functions. The former require specifying the number of 
resources (e.g., number of cores), which is not an input
that serverless users are required to provide. The latter assume
that any server can execute an incoming request, whereas in
serverless computing, specific containers that are already “warm”
(have an active container of that application) are preferred
in order to prevent cold starts. 
On the other hand, the default schedulers built in most widely utilized open-source serverless platforms such as OpenWhisk (https://openwhisk.apache.org/) 
employ locality-based criteria, {\it i.e.,} co-locating invocations
of the same function to a randomly-selected worker without 
taking into consideration load conditions; these techniques are shown to be
ineffective, unable to handle highly-skewed workloads.
Serverless platforms such as OpenFaaS (https://www.openfaas.com/)
treat serverless functions similarly to classic server
workloads and employ auto-scaling when certain pre-defined thresholds
({\it i.e., } based on CPU or memory utilization) are exceeded.
Such coarse-grain policies are {\it reactive} in nature
and thus cannot handle burstiness or highly dynamic serverless workloads, 
and thus lead to high tail latencies and slowdowns.

In this paper we present our approach for rapid container provisioning to support real-world FaaS workloads that exhibit highly dynamic patterns. Our goal is to meet invocation rate and execution time requirements for latency-sensitive big data applications with resource and monetary cost efficiency.
Our work advances state-of-the-art methods as we address the resource provisioning problem even in cases with zero \textit{a priori} knowledge, which has not been exploited by existing works.
In cases that knowledge from past runs is not available to estimate the amount of resources and make load predictions, we utilize a graph similarity approach to capture the similarities among serverless applications, exploiting the graph edit distance metric\cite{zeng2009comparing}, which has exhibited superior performance to alternative techniques.
This is the first work, that we know of, to tackle the problem of predicting the appropriate amount of resources based on the similarity of performance with existing serverless applications with similar codebase. 

In our work we make the following contributions:
\begin{itemize}
    \item We propose SLOPE (Serverless LOad PrEdiction), a framework for estimating the amount of resources required to support different workloads in a serverless environment. 
    \item We build SLOPE by employing a neural network prediction model that allows us to predict efficiently the appropriate number of function replicas as well as select the appropriate configuration to satisfy real-time deadlines, while minimizing operating costs.
    \item We exploit the idea that similar application graphs share certain properties ({\it i.e.,} execution time), 
    and thus we use the appropriate prediction model for applications that exhibit similar graphs
    via a Graph Edit Distance (GED) metric and derive appropriate configurations that satisfy 
    user throughput and application completion time constraints, even for serverless applications \textit{with zero a priori knowledge}.
    \item We have implemented our prototype on top of Apache Mesos and Mesosphere Marathon and evaluated SLOPE using real-world datasets.
    \item Our detailed experimental results illustrate the working and benefits of our approach, which can reduce the operating costs by 66.25\% on average.
\end{itemize}

\section{System Model and Problem Definition}

\subsection{System Model}

{\bf Serverless Model.} 
The Serverless computing model offers a scalable and elastic abstraction where the application code is deployed at the granularity of a function, with a seamless method for autoscaling, using ephemeral containers and can be invoked upon receiving a request. 
The number of active instances
can be specified either by the user or can be adapted dynamically
based on the request rate. During periods of high
load, this number can be adapted automatically
to adjust to the increased traffic, or even decreased to zero during extending periods of inactivity, to keep the total execution cost low.
In this paper, we consider $k$ heterogeneous serverless functions that are hosted on a serverless environment. More formally, let $\mathcal{F}:\{f_1,...,f_k\}$ denote this set of $k$ heterogeneous serverless functions hosted in this environment.

{\bf Containers. } The serverless computing model allows developers to build applications that exploit serverless functions using Docker or custom container images. 
Each function is instantiated in a separate container.
As an application scales out, new container instances are created on-demand. These instances are known as replicas\footnote{We use the terms function replicas and container replicas interchangeably in the paper to denote the number of instances for an application function.} 
and can run in parallel. 
Let $|r_{f_k}^{j}|$ denote the number of replicas for function $f_k$ instantiated in $j$ separate containers, where each $w_j$ container is allocated with $m_{w}$ MBs of memory and $c_{w}$ CPUs. We assume, that, the containers for all replicas of a function $f_k$ are homogeneous, this means that all containers of function $f_k$ have the same CPU and memory allocation. 
The total allocated CPUs and memory for one function is the sum of the CPUs and memory allocated for all replicas of the function. More formally, a container is modelled as follows:
$w_j=\{m_{w},c_{w}\}$
and the configuration of the function $f_k$ (that is, the number of function replicas as well as the memory and CPU allocated for each function replica) is expressed via the following vector
$\overrightarrow{f_k}=\{|r_{f_k}^{j}|,w_j\}$,
where $|r_{f_k}^{j}|$ denotes the number of replicas.

{\bf Function Execution Time. } 
Each function $f_k$ is characterized by the amount of time $\mathcal{T}(f_k)$ it needs to compute. The mean execution time depends on the algorithmic complexity of the application code and the size of the container $w_j$ (memory $m_w$ and CPU $c_w$), hosting the execution environment of the serverless function\cite{perez2018serverless}. Let $\mu^k$ denote the mean execution time of function $f_k$. The mean execution time can be estimated as the ratio of the total execution time of all $N$ requests for function $f_k$ served by the serverless infrastructure over the number of times the serverless function $f_k$ is invoked. More formally,
$\mu^k=\frac{\sum^{N}\mathcal{T}(f_k)}{N}$.

{\bf Workload Completion Time.} SLOPE's goal is to optimize the average Workload Completion Time (WCT) by predicting the amount of resources required to satisfy a certain rate of incoming requests.
The WCT of a serverless function $f_k$ is
defined as the time $\mathcal{T}_{wct}(f_k)$ it takes from the first request arrival until the total number of requests completes. It is the end-to-end time required to serve these requests. The workload completion time consists of: \textit{i) the initialization overhead}, $\mathcal{T}_{init}(f_k)$, that is, the amount of time required to 
instantiate all the containers for the execution of the serverless application,
\textit{ii) the queueing time} in the platform queues $qu_k$, and \textit{iii) execution time required} for the workload to run, that is the ratio of the mean execution time $\mu^k$ times the number $N$ of the function $f_k$ invocations over the number of replicas $|r_{f_k}^{j}|$.
To complete execution by a service level objective(SLO) deadline $d_k$, we require that
$
  \mathcal{T}_{wct}(f_k)=\mathcal{T}_{init}(f_k)+\frac{\mu^k*N}{|r_{f_k}^{j}|} + qu_k  
$ 
and
$\mathcal{T}_{wct}(f_k) \leq d_k$.

\subsection{Problem Definition}
Let a cloud computing platform hosting containerized applications.
SLOPE's objective is to determine 
the appropriate \textit{number} (number of replicas $|r_{f_k}^{j}|$) as well as the \textit{configuration} of containers (memory and CPU size) to spawn 
for the specific function, such that it will minimize the workload completion time $\mathcal{T}_{wct}(f_k)$, and meet its SLO time deadline $d_k$. 
The WCT is a linear function of the initialization time, the execution time for the workload to run $\frac{\mu^k*N}{|r_{f_k}^{j}|}$ and the waiting time in the platform queues. The function initialization time is dependent on the orchestrator, the code size and number of replicas of the application, while, the queueing time is related to the total number of functions scheduled for execution. The parameter that affects mostly the workload completion time and can be further optimized is the execution time for the function to run.

Given a set $\mathcal{W}$ of $i$ possible configurations for a serverless function $f_k$, where
$\mathcal{W} = \{\overrightarrow{f_k}^{1},...,\overrightarrow{f_k}^{i}\} =\{(|r_{f_k}^{j_{1}}|,m_{w_{1}},c_{w_{1}}),...,(|r_{f_k}^{j_{i}}|,m_{w_{i}},c_{w_{i}})\}$ 
the problem is to select the appropriate configuration such that the Workload Completion Time of a function $f_k$, $\mathcal{T}_{wct}(f_k)$, will satisfy a certain SLO deadline. This is related to estimating the number of function instances $|r_{f_k}^{j}|$ to spawn that can run in parallel i.e. maximize the probability $\mathcal{P}$ that a specific configuration and therefore, a specific number of instances 
can fulfil the user constraints and satisfy the SLO constraint. More formally, the problem can be formulated as:
\begin{align}
    max \mathcal{P}(\mathcal{W}=\overrightarrow{f_k}^{i})\\
    s.t. \mathcal{T}_{wct}(f_k) \leq d_k
\end{align}

\section{SLOPE Methodology}

This section highlights the design principles of our resource configuration prediction framework for serverless functions.

\subsection{Resource prediction}
SLOPE utilizes neural networks, 
as an effective approach due to their \textit{design primitive} and \textit{transfer learning capability}. 
The latter characteristic is fully exploited in SLOPE, since the sequential neural network model introduced, can be reused by multiple serverless applications. A Sequential model\cite{donkers2017sequential} is appropriate for a plain stack of layers i.e. it allows us to build a model by stacking layers of nodes (neurons) on top of each other. 
Each argument of the Sequential constructor is a layer of neurons; in this case Dense layers.

{\bf Input Layer.} 
Each neuron has an activation function which computes the value that is passed on to the neurons in the next layer. In SLOPE we choose the ReLU function as an activation function in all layers apart from the last one, which has shown faster convergence times as shown in various works in the bibliography\cite{krizhevsky2012imagenet}, as it requires the estimation of a max value in each neuron rather than the estimation of exponential formulas compared to the sigmoid activation function. 
The input layer takes into consideration the container allocation, {\it i.e.,} the CPUs and memory allocated for the function as well as the request rate of the function.

{\bf Output Layer.} The goal of the neural network is to predict the appropriate number of function replicas required in order to satisfy the specific request rate, as imposed from the developer of the function. The number of predicted necessary instances that will satisfy the user imposed constraints is translated to a label. We utilize the softmax activation function since it normalizes the output of our network to a probability distribution over predicted output classes\cite{luce1977choice}. The softmax activation function implies that we have different probabilities among the different labels.

{\bf Loss Function.} 
We use categorical cross-entropy \cite{zhang2018generalized} (CCE), 
a widely used loss function when optimizing classification models, since using the cross-entropy error instead of the sum-of-squares error function for a classification problem leads to faster training as well as improved generalization of the model\cite{gordon2020uses}. 
There exist also different loss metrics such as the Kullback–Leibler divergence (KLDE) \cite{cao2020deconvolutional} 
and the Poisson distribution (PSSE)\cite{magill2018neural}, 
but are outperformed by CCE loss.

\subsection{Serverless Application Similarity}

Unlike works in the literature \cite{yu2021faasrank}, in this work we use neural networks to learn the most appropriate resource provisioning configuration for serverless applications in order to satisfy certain service level objective deadlines. 
Reinforcement learning models have been shown to be a good fit \cite{yu2021faasrank} for learning policies for computer systems, because the model agents are capable of learning from real-world workloads and operating conditions without human-designed inaccurate assumptions and interference, accumulating knowledge from previous experience \cite{kornblith2019similarity}.  
Consistent to earlier works \cite{zacheilas2018dione,xin2022locat} that exploit execution plans, we make the assumption that serverless applications with similar codebase will have the same behavior for the same size of input (thus it allows us to estimate faster the resource provisioning configurations).

{\bf Estimating the execution times of serverless applications with zero \textit{a priori} knowledge.} To overcome the limitations regarding the size of the trained model and retraining the neural network for each serverless application, we propose a different approach utilizing previously computed prediction models for estimating the execution times of applications that perform similar processing, given that models for a different but somehow similar problem can be reused partly or entirely to accelerate the training of other similar models.
SLOPE exploits the notion of call graphs\cite{obetz2019static,kannan2019grandslam} 
for estimating the similarity between serverless applications. 
We exploit the fact that different applications may consist of similar call graphs and thus the similarity in the functions leads also to similar execution times. 
To compute the similarity between the call graphs of the serverless applications we need a graph similarity measure. There are two well known graph similarity metrics, the graph edit distance (GED) \cite{zeng2009comparing,zacheilas2018dione} and the maximum common subgraph (MCS) \cite{bunke1998graph}.
We decided to opt for the Graph Edit Distance metric as it has been accepted as the most appropriate measure for representing the distance between graphs. GED defines the similarity between two graphs by the minimum amount of required distortions to transform one graph into the other, is error-tolerant and can identify similar graphs even in the presence of noise and errors.

{\bf Defining the dissimilarity score.} We exploit a novel metric \cite{zacheilas2018dione}, called dissimilarity score to capture the degree of
similarity between the call graphs of two serverless applications. In SLOPE, this metric depicts the minimum required distortions to make the call graph of one serverless applications identical to the
call graph of the other applications. The lower the value
of this score, the more similar are the two applications.
More formally, assume that $DS(CG(f_{k1}), CG(f_{k2}))$ represents
the Dissimilarity Score (DS) between two call graphs, $CG(f_{k1})$
and $CG(f_{k2})$ of the serverless applications consisting of functions $f_{k1},f_{k2}$ respectively and that $ged_{CG(f_{k1}),CG(f_{k2})}$ is the GED distance between call graph $CG(f_{k1})$ and call graph $CG(f_{k2})$. The main idea is to match the call graph $CG(f_{k1})$ with exactly the call graph of $CG(f_{k2})$, 
compute their GEDs (i.e., the number of
necessary distortions to make two call graphs identical) and
then aggregate these values to compute the DS metric.

{\bf GED Computation.} Thus, in order to
compute the $DS(CG(f_{k1}), CG(f_{k2}))$ it is necessary to compute
the GED distance between the call graphs of the two serverless applications. 
Computing GED is an NP-Hard problem and for this reason we decided to use a well-known approximation technique\cite{zeng2009comparing} that is able
to effectively and in polynomial time approximate the GED between two call graphs, by transforming them to multisets of star structures.

{\bf Detecting the most similar application.} The computation of the $DS$ of two call graphs allows to detect whether a serverless function exists for which we have already built a prediction model. 
We find the serverless application that leads to the minimum $DS$ and then examine whether this value
is smaller than a pre-defined threshold $\mathcal{T}$, which can be tuned dynamically based on the degree of similarity we target.
If this condition is true we simply return the already built prediction model.
In the case that the minimum $DS$ score is larger than T then we proceed with the next most similar serverless application in the $HQ$ set.
Thus, we can estimate the number of instances to \textit{prewarm} for execution in order to meet the developer SLO deadline, even if there is no prior knowledge regarding the application resource needs or performance.

\section{Implementation}\label{sec:implement}

\begin{figure}[t!]\centering
\begin{minipage}{\linewidth}\centering
\includegraphics[width=\linewidth]{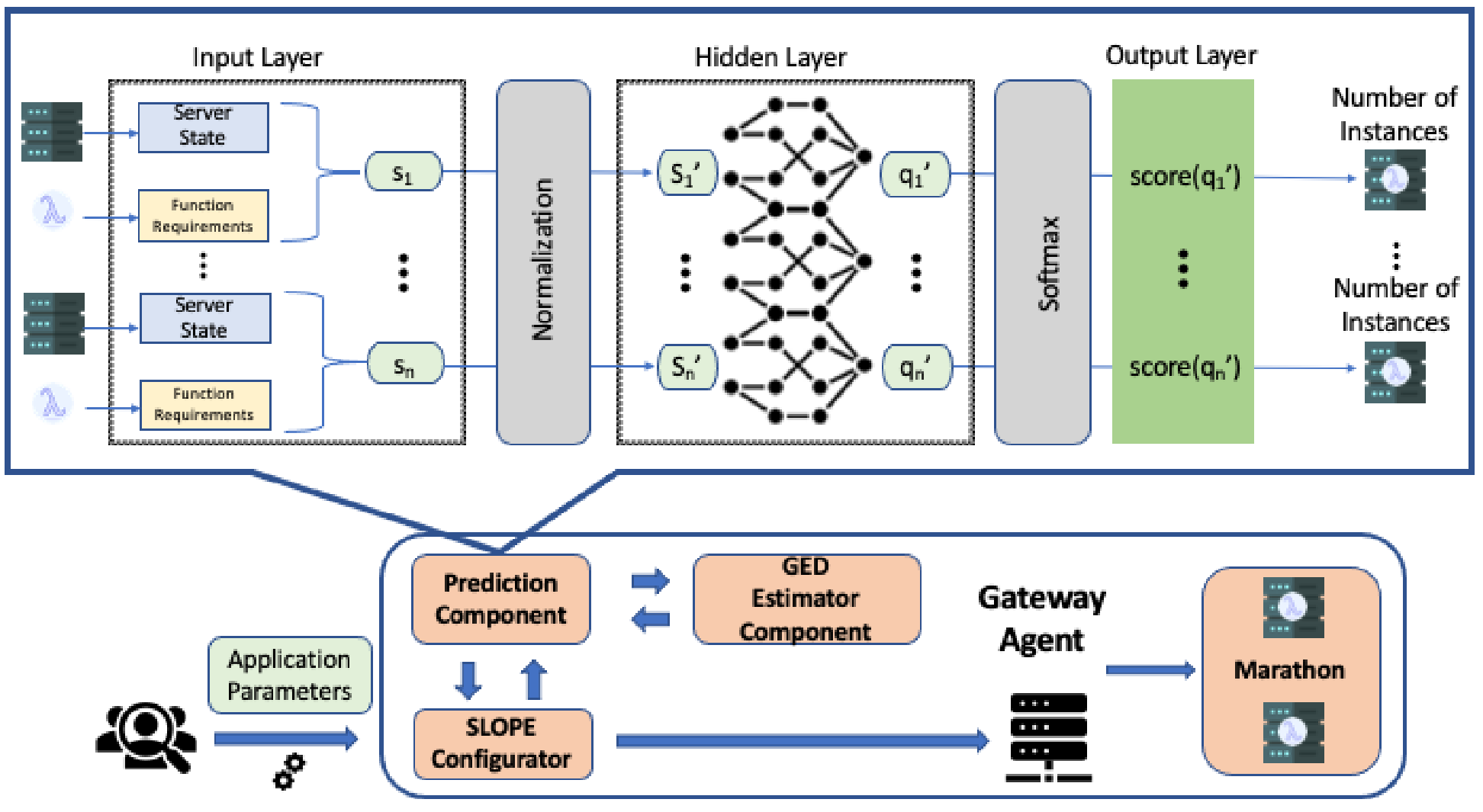}
\caption{SLOPE architecture}\label{fig:overview}
\end{minipage}
\end{figure}

{\bf SLOPE Configurator:} 
SLOPE utilizes a frontend, which is in line with state-of-the-art approaches\cite{jiang2021towards} and allows developers to specify their serverless application service level objective deadlines, as well as, upload their code to be executed in the serverless environment. The SLOPE Configurator interacts with the Prediction Component in order to specify the number of instances required for the specific application code to execute and meet the defined SLO deadline and then forwards the estimated configuration and the developers code to be deployed through Marathon.

{\bf Prediction Component:} The Prediction Component is responsible for our neural network model, as well as, 
to interact with the Graph-Edit Distance Estimator Component. 
It estimates the number of instances required in order to satisfy the developer's SLO deadline. 

\begin{figure*}[t!]\centering
\begin{minipage}{0.24\linewidth}\centering
\includegraphics[width=\linewidth]{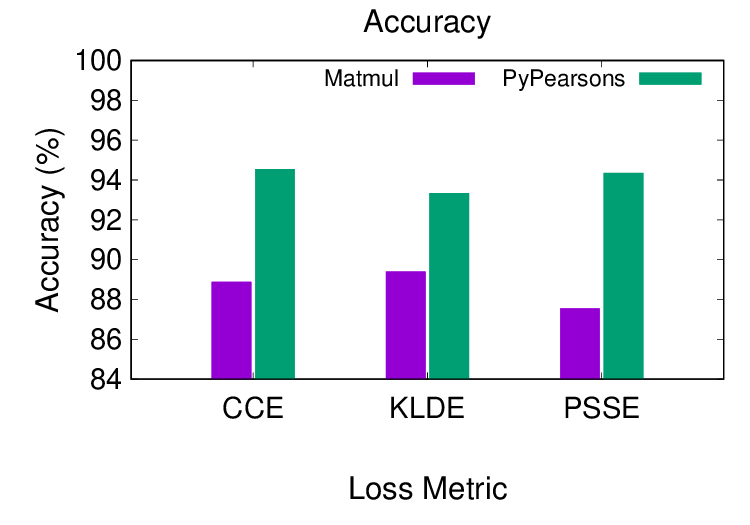}
\caption{Accuracy}\label{fig:accuracy}
\end{minipage}\hfill
\begin{minipage}{0.24\linewidth}\centering
\includegraphics[width=\linewidth]{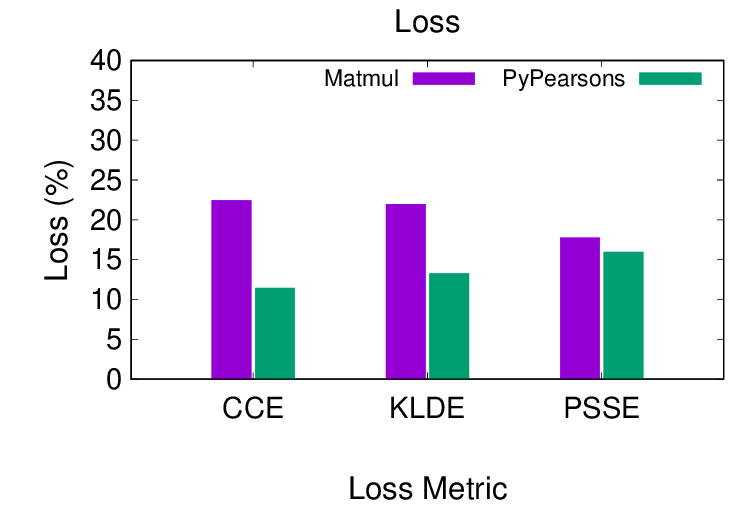}
\caption{Loss}\label{fig:loss}
\end{minipage}\hfill
\begin{minipage}{0.24\linewidth}\centering
\includegraphics[width=\linewidth]{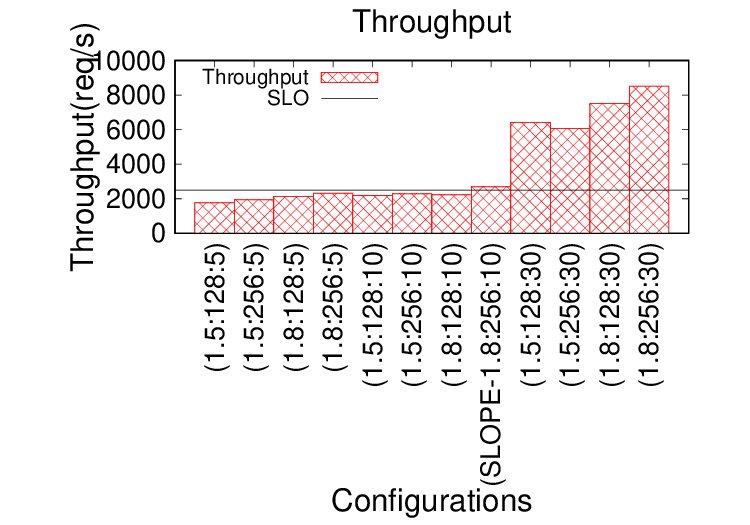}
\caption{Matmul throughput}\label{fig:matthrough}
\end{minipage}\hfill
\begin{minipage}{0.24\linewidth}\centering
\includegraphics[width=\linewidth]{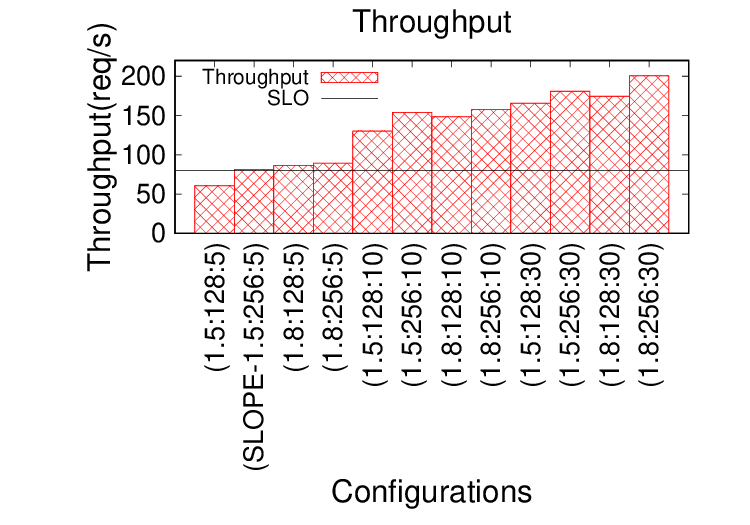}
\caption{PyPearsons throughput}\label{fig:pearsthrough}
\end{minipage}\hfill
\end{figure*}

{\bf Graph-Edit Distance Estimator:} We built the Graph-Edit Distance Estimator Component to exploit the similar performance of similar serverless applications based on their call-graphs. It is responsible to run the calculation of the $DS$ dissimilarity score function, defined in the previous section, in order to quantify the similarity between existing function models in the Prediction Component and the developer's submitted code.

{\bf Gateway Agent:} We developed a Gateway as in \cite{tsenos2022amesos} to deploy a new function and to scale up/down existing ones. It also acts as a Proxy for function invocations, which are propagated to the deployed function containers. 

{\bf Marathon:} We use Mesosphere Marathon (https://mesosphere.github.io/marathon/) to start serverless function containers on our Apache Mesos cluster. Mesos abstracts the compute resources away from machines (physical or virtual).
Each container listens to port 8080 and Mesos maps that port to a random port of the host Agent. Each function can be invoked by receiving an HTTP POST request. An implementation overview is presented in Figure \ref{fig:overview}.

\section{Experimental Evaluation}

\subsection{Experimental Setup}
Similar to related works \cite{zacheilas2018dione,xin2022locat,kannan2019grandslam}, we conducted our experiments in our local cluster comprising 7 nodes (Intel i7-7700 3.6GHz processors), with 56 CPUs and 112GB of RAM available in total, running on Ubuntu 20.04 LTS, with 1Gbps Ethernet. 
We run Apache Mesos 1.9 as our serverless platform and we use Marathon 1.5 in order to deploy Docker containers on top of Mesos. 
Due to limitations imposed by our serverless infrastructure, we varied the number of serverless function instances from 5 up to 30. 
We used OpenFaas Python templates to create our functions and PyCG\cite{salis2021pycg} to extract the call graphs for Python applications.

\subsection{Serverless Functions Benchmark}
We evaluated the performance of our approach using real world application scenarios from state-of-the-art performance benchmarks\cite{kim2019functionbench,jiang2021towards,zacheilas2018dione,muller2020lambada}.
Below, we give a brief description for each one of those application functions ({\bf AF}):

\noindent\textit{{\bf AF1} - Matmul:} Matmul performs square matrix multiplications. It is considered as CPU benchmark\cite{malawski2017benchmarking,manner2021optimizing}, which is mainly used to measure the CPU-bound performance.

\noindent\textit{{\bf AF2} - PyPearsons:} PyPearsons is a Python implementation of Pearsons correlation over a smart-city sensor network and for a given set of geospatial coordinates it returns a list of the most correlated sensors.

\subsection{Prediction Performance}

{\bf Dataset:} We evaluated the prediction performance of our neural network using the aforementioned workloads, with four different types of memory and CPU configurations, and also varied the number of replica instances. We performed 1000 runs using $Hey$ (https://github.com/rakyll/hey), for every possible configuration, and aggregated these results in order to construct the training dataset for SLOPE's neural network.

{\bf Prediction Metrics:} We evaluate the prediction performance using metrics similar to existing works in the literature.

\noindent{\bf Accuracy.} The accuracy metric illustrates the frequency with which predicted labels match the true labels.

\noindent{\bf Loss.} 
We evaluated our framework with different available types of losses in the literature to find the appropriate for our problem, such as CCE \cite{zhang2018generalized}, KLDE \cite{cao2020deconvolutional} and PSSE \cite{magill2018neural}.

{\bf Findings: }
In Figures \ref{fig:accuracy} and \ref{fig:loss}, we illustrate the behaviour of each different serverless function, where in the x-axis we vary the loss metric utilized for the training of the neural model. The results show that each different function can achieve very good performance in terms of accuracy and loss, when using the categorical cross entropy as the loss metric required for training. We reason these results due to the fact that KLDE calculates the relative entropy between two probability distributions, whereas cross-entropy can be used to calculate the total entropy between two distributions.

\subsection{Serverless performance}

{\bf Performance Metrics:} We utilize the \textit{throughput} as the performance metric of our framework \cite{kim2020automated}, which is defined as the number of requests/second that are successfully served from the serverless system.

{\bf Findings: }
In Figures \ref{fig:matthrough} and \ref{fig:pearsthrough}, we draw the throughput achieved by each one of the configurations for a given SLO (equal to 2500 requests / second for matmul and 80 requests / second for PyPearsons (which is computationally intensive)). In the x-axis, we draw all the examined configurations, in y-axis we draw the number of throughput achieved by each one of the examined configuration, as well as the one predicted by SLOPE. Compared to choosing naively the greatest number of replicas (i.e. 30) and the largest configuration available, SLOPE can save up to 68.32\% for the Matmul serverless app in terms of operations costs \textit{without overprovisioning}.

\section{Related Work}

{\bf Container Optimization: } The authors of \cite{silva2020prebaking} focus on how to optimize the container creation by using shortcuts based on checkpoint-and-restore procedures, without the need of recreating the docker container image. In \cite{hunhoff2020proactive}, the authors propose an approach where the developers can specify functionality to perform before a given function executes.
In \cite{oakes2018sock}, they improve the container boot process to achieve cold starts in the low hundreds of milliseconds.
In \cite{du2020catalyzer} the authors demonstrate snapshot and restore in VMs and unikernels to achieve millisecond serverless cold starts. The work of \cite{shillaker2020faasm} proposes a new lightweight isolation mechanism which restores Faaslets from already-initialised snapshots.

{\bf Prediction methods: }
In \cite{shahrad2020serverless}, they propose an adaptive resource management policy, but it does not consider the similarity of serverless applications as we do in our work.
The work of \cite{fuerst2021faascache} proposes a keep-alive policy for the OpenWhisk serverless platform using a function hit-ratio curve for determining the percentage of warm-starts at different server memory sizes. In \cite{kannan2019grandslam}, the authors estimate the completion time of requests that propagate through a set of individual microservices, but they do not focus on identifying the best resource provisioning configuration as we do in our work. In \cite{fu2019edgewise}, they incorporate a congestion-aware scheduler into an edge streaming process environment, but can not be applied to our setting, in which we focus on batches of requests.

\section{Conclusions}

In this paper, we presented SLOPE, a framework for estimating the amount of resources required to support different workloads in a serverless environment, using neural networks. 
We exploited the graph edit distance metric to identify similarities between similarly behaving serverless applications and derive the appropriate configuration to satisfy certain SLO constraints, 
even in cases  of serverless applications \textit{with zero a priori knowledge}. Finally, we presented our prototype on top of Apache Mesos and Marathon and evaluated its efficiency using real datasets achieving a reduction of the operating costs by up to 66.25\% on average.

\section*{Acknowledgment}
This research has been financed by the European Union through the EU ICT-48 2020 project TAILOR (No. 952215), the H2020 AutoFair project (No. 101070568) \& the Horizon Europe CoDiet project (No. 101084642) \& the Hellenic Foundation for Research and Innovation (HFRI) under the 3rd Call for HFRI PhD Fellowships (Fellowship Number: 6812).

\bibliographystyle{IEEEtran}

\bibliography{biblio}

\end{document}